\documentclass[11pt]{article}
\usepackage{amssymb,latexsym,amsmath}
\usepackage[dvips]{graphicx}
\usepackage{cite}
\newtheorem{theo}{Theorem}

\newtheorem{lemm}[theo]{Lemma}

\def\nn{\nonumber}

\def\diag{\mathop{\rm diag}\nolimits}
\def\qdots{\mathinner{\mkern1mu\raise1pt\vbox{\kern7pt\hbox{.}}\mkern2mu
 \raise4pt\hbox{.}\mkern2mu\raise7pt\hbox{.}\mkern1mu}}
\def\Z{{\mathbb Z}}
\def\N{{\mathbb N}}

\newcommand{\hatH}{{\hat H}}
\renewcommand{\atop}[2]{\genfrac{}{}{0pt}{}{#1}{#2}}

\begin{document}
\begin{center}
{\Large \bf
Quantum communication through a spin chain\\[2mm]
with interaction determined by a Jacobi matrix
}\\[5mm]
{\bf R.~Chakrabarti\footnote{E-mail: ranabir@imsc.res.in; Permanent address: 
Department of Theoretical Physics, University of Madras, Guindy Campus, Chennai 600 025,
India} and J.\ Van der Jeugt}\footnote{E-mail:
Joris.VanderJeugt@UGent.be}\\[1mm]
Department of Applied Mathematics and Computer Science,
Ghent University,\\
Krijgslaan 281-S9, B-9000 Gent, Belgium.
\end{center}

\vskip 10mm
\noindent
Short title: correlation functions for quantum communication



\begin{abstract}

We obtain the time-dependent correlation function describing the evolution of a
single spin excitation state in a linear spin chain with isotropic 
nearest-neighbour $XY$ coupling, where the Hamiltonian is related to the Jacobi
matrix of a set of orthogonal polynomials. For the Krawtchouk polynomial case 
an arbitrary element of the correlation function is expressed in a simple 
closed form. Its asymptotic limit corresponds to the Jacobi matrix of the 
Charlier polynomial, and may be understood as a unitary evolution resulting 
from a Heisenberg group element. Correlation functions for Hamiltonians 
corresponding to Jacobi matrices for the Hahn, dual Hahn and Racah polynomials 
are also studied. For the Hahn polynomials we obtain the general correlation 
function, some of its special cases, and the limit related to the Meixner 
polynomials, where the $su(1,1)$ algebra describes the underlying symmetry.
For the cases of dual Hahn and Racah polynomials the general 
expressions of the correlation functions contain summations which are not of 
hypergeometric type. Simplifications, however, occur in special cases.   

\end{abstract}
\section{Introduction}
Transfer of a known or an unknown quantum state from one site to another is a 
key requirement in linking neighbouring small quantum processors for
facilitating large scale quantum computation. S.~Bose~\cite{Bose2003,Bose2005} 
introduced linear spin chains as a channel for such short distance quantum 
communication. Such a connector has an inherent advantage as it renders the 
quantum processor and the communicating channel to be made of the same 
physical system. This
eliminates the need of developing interfaces. Transmission of data in such 
linear quantum registers has been the subject of many 
investigations~\cite{Bose2003,Christandl2004,Christandl2005,PPKF05,Kay2009}. 
Using the classical concept of group velocity of a wave packet it has been 
observed~\cite{OL04} that one-dimensional spin rings allow 
high-fidelity transmission of quantum states if both communicating parties have
access to a finitely limited number of qubits in the ring. 
Quantum communication with 
closed spin ring with twisted boundary conditions 
has been discussed in Ref.~\cite{Bose2005}. An excellent review~\cite{Bose2007}
and references therein describe the current developments in this field. 

The transmission of quantum states can in principle be performed by a chain of qubits coupled via
the Heisenberg or the $XY$ interactions~\cite{DiVicenzo2000,Benjamin2002,Zhou2002,Benjamin2003}. Interesting situations arise if one assumes to have individual control of the 
nearest-neighbour couplings in the spin chain.
The idea of pre-engineered interqubit couplings has been discussed considerably~\cite{Yung2005,Karbach2005}.
One of the advantages of well-chosen controlled couplings is that one can obtain mirror inversion of 
a quantum state with respect to the center of the chain, and that perfect transfer of quantum states
is possible~\cite{Albanese2004,Christandl2004,Kay2009} at certain specified times over arbitrary length of the spin chain. Propagation of entangled states in anisotropic $XY$ spin chains has been discussed in~\cite{AOPFP04}. 
Using a system 
based on a dispersive qubit-boson interaction to mimic $XY$ coupling that 
relaxes the nearest-neighbour restriction the transfer fidelity of the chain 
has been found~\cite{PPKF05} to achieve a nearly optimal value.     

One of the main results of~\cite{Albanese2004} is the introduction of two (analytic) mirror-periodic
Hamiltonians that allow for perfect transfer. To achieve perfect mirror 
inversion of an arbitrary many excitation state it is sufficient to consider 
transformations of all single excitation states to their mirror images as the 
time-dependent transition amplitudes for multiple excitation states may be 
constructed~\cite{Lieb1968, Albanese2004} via the Slater determinant of their 
single excitation counterparts. The eigenstates of the single excitation 
sectors of these Hamiltonians are related~\cite{Albanese2004} to discrete orthogonal polynomials, 
namely Krawtchouk polynomials and dual Hahn polynomials.

In the present paper we investigate such systems from a general point of view.
We shall assume that the chain of qubits is described by a Hamiltonian of $XY$ type, in such a way that the
interaction strengths (and qubit energies) are related to the Jacobi matrix of a system of
discrete orthogonal polynomials.
This will allow us to give a general expression for the transition amplitude 
of a single excitation
from the $s$th site (sending site) to the $r$th site (receiving site) in the chain of spins.
For some types of discrete orthogonal polynomials, the expression of these transition amplitudes can be 
simplified, leading to some interesting properties.

Let us consider a by now classical system of $N+1$ interacting qubits (spin $1/2$ particles) in
a quantum register, with an isotropic Hamiltonian of $XY$ type:
\begin{equation}
\hatH=\frac12 \sum_{k=0}^{N-1} J_k(\sigma^x_k\cdot\sigma^x_{k+1}+\sigma^y_{k+1}
\cdot\sigma^y_{k}) + 
\frac12 \sum_{k=0}^N h_k (\sigma^z_k+1),
\label{Ham1}
\end{equation}
where $J_k$ is the coupling strength between the qubits located at sites $k$ and $k+1$,
and $h_k$ is the ``Zeeman'' energy of a qubit at site~$k$. So the subindex $k$ ($k=0,1,2,\ldots,N$) labels
the position of the qubit in the chain, and the superindex refers to the Pauli matrices
$\sigma^x$, $\sigma^y$ and $\sigma^z$. The Hamiltonian~\eqref{Ham1} preserves 
the total spin: $[\hatH,\;\sum_{k=0}^N\,\sigma^z_k] = 0,$ and, therefore, we may 
analyze various spin excitation sectors separately.

To describe the Hilbert space associated with the Hamiltonian, one adopts a standard
fermionization technique~\cite{Lieb1968}. The Jordan-Wigner transformation~\cite{Jordan1928} maps the Pauli matrices to spinless lattice fermions 
$\{a_{k}, a_{k}^{\dagger}|\;k = 0, 1, \ldots, N\}$ obeying the 
anticommutation rules 
\begin{equation}
\{a_{k}^{\dagger}, a_{\ell}\} = \delta_{k, \ell},\qquad 
\{a_{k}, a_{\ell}\} = \{a_{k}^\dagger, a_{\ell}^\dagger\} =0 \qquad \forall k, \ell \in \{0, 1, \cdots, N\}.
\label{anticomm}
\end{equation} 
We may now recast the Hamiltonian~\eqref{Ham1} as
\begin{equation}
\hatH= \sum_{k=0}^{N-1} J_k(a_k^\dagger a_{k+1}+a_{k+1}^\dagger a_k) + \sum_{k=0}^N h_k a^\dagger_k a_k
\label{Ham2}
\end{equation}
that describes a set of $N+1$ fermions on a chain with 
nearest-neighbour interaction (hopping between adjacent sites of the chain), and subject to a non-uniform background
magnetic field denoted by $h_k$ ($k=0,1,\ldots,N$).
We shall assume that the system is initially in its completely polarized  
ground state $|{\bf 0}\rangle=|00\cdots0\rangle=|0\rangle \otimes
|0\rangle\otimes\cdots\otimes|0\rangle$, where $|0\rangle$ denotes the spin down state. 
Let $|k)=|00\cdots 010\cdots 0\rangle=a_k^\dagger|{\bf 0}\rangle$ ($k=0,1,\ldots,N$) 
denote a state in which there is a single fermion at the site~$k$
and all other sites are empty, i.e.\ $|k)$ describes the state in which the spin at the
site~$k$ has been flipped to $|1\rangle$. 
Clearly, the set of states $|k)$ ($k=0,1,\ldots,N)$ forms a basis for the single-fermion states of
the system, and we may represent them by the standard unit vectors in column matrix form:
\begin{equation}
|k) = \left( \begin{array}{c} 0\\0\\ \vdots \\ 1 \\ \vdots \\0\end{array}\right) 
\qquad (k=0,1,\ldots,N).
\end{equation}
In this single-fermion basis, the Hamiltonian $\hatH$ takes the matrix form
\begin{equation}
M=\left(
\begin{array}{ccccc}
h_0 & J_0& 0 & \cdots & 0 \\
J_0 & h_1 & J_1 & \cdots & 0\\
0 & J_1 & h_2 & \ddots &  \\
\vdots & \vdots & \ddots & \ddots& J_{N-1}\\
0 & 0 &  & J_{N-1} & h_N
\end{array}
\right).
\label{Ham-M}
\end{equation}
The dynamics (time evolution) of the system is completely determined by the eigenvalues $\epsilon_j$
and eigenvectors $\phi_j$ of this matrix. 
It is then, as noted before, a standard technique~\cite{Lieb1968,Albanese2004} 
to describe the $n$-fermion eigenstates of $\hatH$ ($n\leq N$) using the
single-fermion eigenstates $\phi_j$ and Slater determinants.
For this reason we concentrate here on the single-fermion eigenstates.

The matrix $M$ in~\eqref{Ham-M} is real and symmetric, so the spectral theorem~\cite{Golub1996}
implies that it can be written as
\begin{equation}
M=UDU^T
\label{M=UDU}
\end{equation}
where $D$ is a diagonal matrix and $U$ an orthogonal matrix:
\begin{align}
& D= \diag (\epsilon_0,\epsilon_1,\ldots, \epsilon_N),\\
& U U^T=U^TU=I.
\end{align}
The entries of $D$ are the single-fermion energy eigenvalues, and the columns of the matrix $U$
are the (orthonormal) eigenvectors of $M$, i.e.\ the single-fermion eigenstates:
\begin{equation}
\phi_j= 
\left( \begin{array}{c} U_{0j} \\ U_{1j} \\ \vdots \\ U_{Nj} \end{array}\right) =
\sum_{k=0}^N U_{kj}\;|k) = \sum_{k=0}^N U_{kj}\;a_k^\dagger |{\bf 0}\rangle \qquad(j=0,1,\ldots,N),
\end{equation}
with $\hatH\phi_j = M\phi_j = \epsilon_j\,\phi_j$.
{}From the orthogonality of $U$, the inverse relation follows:
\begin{equation}
|k) = \sum_{j=0}^N U_{kj} \phi_j.
\label{inverse}
\end{equation}

We now turn to the dynamics of the system under consideration, 
described by the unitary time evolution operator 
\begin{equation}
\mathcal{U}(t) \equiv \exp(-it\hatH).
\label{evolution}
\end{equation}
Assume that the ``state sender'' is located at site $s$ of the spin chain, and the ``state receiver''
at site $r$ ($s$ and $r$ are site labels, belonging to $\{0,1,\ldots,N\}$). 
At time $t=0$ the sender turns the system into the single spin state $|s)$.
After a certain time~$t$, the system evolves to the state $\mathcal{U}(t)|s)$ 
that may be expressed as a linear superposition of all the single spin states.
So the transition amplitude of an excitation from site $s$ to site $r$ of the 
spin chain is given by the time-dependent correlation function
\begin{equation}
f_{r,s}(t) = (r|\mathcal{U}(t)|s).
\label{frs}
\end{equation}
This is the central quantity of this paper (sometimes referred to as the
``correlation function''), and the square of its modulus 
gives the transition probability from the $s$th to the $r$th spin 
excitation state.
Note that it can be expressed by means of the orthogonal matrix $U$ appearing in~\eqref{M=UDU}.
Indeed, using the definition~\eqref{frs}, the expansion~\eqref{inverse}, and 
the orthogonality of the states $\phi_j$, one finds:
\begin{align}
f_{r,s}(t) &= 
 \langle \sum_{k=0}^N U_{rk}\phi_k | \exp(-it\hatH) \sum_{j=0}^N U_{sj}\phi_j\rangle\nn\\
&=  \langle \sum_{k=0}^N U_{rk}\phi_k | \sum_{j=0}^N U_{sj} e^{-it\epsilon_j}\phi_j\rangle\nn\\
&= \sum_{j=0}^N U_{rj}U_{sj} e^{-it\epsilon_j}.
\end{align}
In other words, using the abbreviation $z=e^{-it}$, one has
\begin{equation}
f_{r,s}(t) = \sum_{j=0}^N U_{rj}U_{sj} z^{\epsilon_j} \qquad(z=e^{-it}).
\label{frsz}
\end{equation}
The purpose of this paper is to show that various interesting closed form expressions
can be given for this crucial quantity $f_{r,s}(t)$, in the case that the fixed values
characterizing the system (the values $J_k$ and $h_k$) are related to the Jacobi matrix
of a set of discrete orthogonal polynomials. 
We shall illustrate this first by means of an example, where the polynomials are Krawtchouk
polynomials. From this example, the general technique will be clear. Then we continue
analyzing some systems related to other classes of orthogonal polynomials.

{}From the outset it is assumed that such chains of qubits can be pre-engineered for any
given set of values $J_k$ and $h_k$, i.e.\ the strength of the couplings can be engineered and
the external static potential at site $k$ is controlled. Such a physical control has
been the subject of many papers (see~\cite{Christandl2004} and references therein).

We end this section by two remarks. 
First of all, the strengths $J_k$ in~\eqref{Ham2}, and thus the off-diagonal elements of $M$ in~\eqref{Ham-M},
are positive. From the mathematical point of view, however, the problem can just as well be
solved for a matrix with negative off-diagonal elements. Indeed, if
\begin{equation}
M'=\left(
\begin{array}{ccccc}
h_0 & -J_0& 0 & \cdots & 0 \\
-J_0 & h_1 & -J_1 & \cdots & 0\\
0 & -J_1 & h_2 & \ddots &  \\
\vdots & \vdots & \ddots & \ddots& -J_{N-1}\\
0 & 0 &  & -J_{N-1} & h_N
\end{array}
\right),
\label{Ham-M'}
\end{equation}
then $M'_{jk}=(-1)^{j+k}M_{jk}$. This implies that $M'$ has the same eigenvalues $\epsilon_j$ as $M$. 
Moreover, the orthogonal matrix $U'$ with $U'_{jk}=(-1)^{j+k}U_{jk}$ diagonalizes $M'$ in the same way
as in~\eqref{M=UDU}, so the eigenvectors of $M'$ (i.e.\ the columns of $U'$) are the same as those of
$M$ up to sign changes in the components. Then the transition amplitude follows from~\eqref{frsz} and
is essentially the same as that corresponding to~$M$: $f'_{r,s}(t)= (-1)^{r+s} f_{r,s}(t)$.
Secondly, also a matrix that differs from the original one by a constant factor and a multiple of the 
identity matrix leads essentially to the same computation. Indeed, let $M'=\lambda M+ \mu I$, where
$\lambda$ and $\mu$ are constants and $I$ is the identity matrix. Then the same matrix $U$ from~\eqref{M=UDU}
diagonalizes $M'$, the only difference being the eigenvalues given now by $\lambda\epsilon_j+\mu$. 
Also, it follows immediately from~\eqref{frsz} that $f'_{r,s}(t)= e^{-it\mu} f_{r,s}(\lambda t)$.

\section{The Jacobi matrix of Krawtchouk polynomials} 
\label{Krawtchouk}
\subsection{Computation of the general correlation function}

Let us start this section by introducing some standard notation and known facts of Krawtchouk polynomials.
We follow the notation of~\cite{Koekoek}; other standard works on (discrete) orthogonal polynomials
are~\cite{Suslov,Ismail}.
The Krawtchouk polynomial of degree $n$ ($n=0,1,\ldots,N$) in the variable $x$, with parameter $0<p<1$ is given by
\begin{equation}
K_n(x)\equiv K_n(x; p,N) = \mbox{$_2F_1$} \left( \atop{-x,-n}{-N} ; \frac{1}{p} \right).
\label{defK}
\end{equation}
The function ${}_2F_1$ is the classical hypergeometric series~\cite{Bailey,Slater}, 
and in this case it is a terminating
series because of the appearance of the negative integer $-n$ as a numerator parameter.
Krawtchouk polynomials satisfy a (discrete) orthogonality relation~\cite{Koekoek}:
\begin{equation}
\sum_{x=0}^N w(x) K_n(x) K_m(x) = d_n \delta_{mn},
\label{orth-K}
\end{equation} 
where $w(x)$ is the weight function in $x$ and $d_n$ is a function depending on $n$:
\begin{equation}
w(x) = \binom{N}{x} \, p^x \, (1-p)^{N-x}\qquad (x=0,1,\ldots,N); \qquad\qquad
d_n = \frac{1}{\binom{N}{n}} \left( \frac{1-p}{p} \right)^n.
\end{equation}
They also satisfy the following three-term recurrence relation:
\begin{equation}
-x K_n(x)  =  n(1-p) \, K_{n-1}(x)- \bigl[ p(N-n)+n(1-p) \bigr] \, K_n(x)
+ \, p(N-n) \, K_{n+1}(x). 
\label{kraw_rec}
\end{equation}

It is often convenient to introduce orthonormal Krawtchouk functions by
\begin{equation}
\tilde K_n(x) \equiv \frac{\sqrt{w(x)} K_n(x)}{\sqrt{d_n}}.
\end{equation}
Then, rewriting the orthogonality relation and the recurrence relation in terms of
the functions $\tilde K_n(x)$, it is easy to obtain the following:
\begin{lemm}[see~\cite{Regniers2009}]
Let $M_K$ be the tridiagonal $(N+1)\times(N+1)$-matrix (Jacobi matrix)
\begin{equation}
\label{MK}
M_K= \left( \begin{array}{ccccc}
             h_0 & -J_0  &    0   &        &      \\
            -J_0 &  h_1  &  -J_1  & \ddots &      \\
              0  & -J_1  &   h_2  & \ddots &  0   \\
                 &\ddots & \ddots & \ddots & -J_{N-1} \\
                 &       &    0   &  -J_{N-1}  &  h_N
          \end{array} \right)
\end{equation}
where
\begin{equation} 
J_n = \sqrt{p(1-p)} \sqrt{(n+1)(N-n)}, \qquad h_n = Np + (1-2p)n,
\label{Jh-K}
\end{equation}
and let $U$ be the $(N+1)\times(N+1)$-matrix with elements $U_{jk}=\tilde K_j(k)$.
Then 
\begin{equation}
U U^T = U^TU=I \qquad\hbox{and}\qquad M_K=UDU^T
\end{equation}
where
\begin{equation}
D= \diag (0,1,2,\ldots,N).
\end{equation}
\end{lemm}
In other words, the eigenvectors of the Hamiltonian (in the single-fermion case) corresponding
to the quantities~\eqref{Jh-K} have components
equal to normalized Krawtchouk polynomials, and the corresponding energy eigenvalues are $\epsilon_j=j$
($j=0,1,\ldots,N$). 
Note that Krawtchouk polynomials have been used before as a basis for quantum chains~\cite{Atakishiev}.
Here, only their evaluations at integer values of the support are used as matrix elements of $U$ 
in the diagonalization process.

Let us now consider the transition amplitude or correlation function:
\begin{align}
f_{r,s}(t) &= \sum_{k=0}^N U_{rk}U_{sk}z^{\epsilon_k} = \sum_{k=0}^N \tilde K_r(k) \tilde K_s(k) z^k \nn\\
 &= \frac{1}{\sqrt{d_r d_s}} \sum_{k=0}^N w(k) K_r(k) K_s(k) z^k \qquad (z=e^{-it}).
\label{sumK}
\end{align}
So we need to compute the quantity in~\eqref{sumK}. 
First of all, note that in general $f_{r,s}(t)$ is a periodic function of $t$ since $z=e^{-it}$.
In particular, it follows from~\eqref{sumK} and the orthogonality relation~\eqref{orth-K}
that $f_{r,s}(t)=\delta_{rs}$ for $t=0$ and for any multiple of $2\pi$. So after a time span of $2\pi$,
the system is back in its original state where only the spin at the sending site $s$ is flipped.

The purpose is now to compute~\eqref{sumK} explicitly. We shall do this in two ways: a classical method
and a group theoretical method. 
The classical method is short and straightforward. Rewriting the polynomials in~\eqref{sumK} as
${}_2F_1$-series, this sum reduces to a classical summation formula given for example 
in~\cite[p.\ 84, (8)]{Bateman}. This leads immediately to the following closed form expression:
\begin{equation}
f_{r,s}(t)= \sqrt{\binom{N}{r}\binom{N}{s}}(\sqrt{p(1-p)})^{r+s} (1-z)^{r+s}(1-p+pz)^{N-r-s}{\ }_2F_1\left(
 \atop{-r,-s}{-N} ; \frac{-z}{p(1-p)(1-z)^2} \right).
\label{frsK} 
\end{equation}

Before discussing some special and interesting cases of this formula, we shall also deduce 
this correlation function in a different way.

\subsection{Group-theoretical computation}

The group-theoretical way to obtain~\eqref{frsK} is somewhat longer, but it does not use any reference
to orthogonal polynomials or summation formulas of hypergeometric type. 
So it sheds another light on why the final formula~\eqref{frsK} is so simple.

Consider the Lie algebra $su(2)$ of quantum angular momentum theory~\cite{Edmonds}, with basis $L_0$, $L_\pm$ and 
commutation relations
\begin{equation}
[L_0,L_\pm]=\pm L_\pm, \qquad [L_+,L_-]=2L_0.
\label{sl2}
\end{equation}
For $N$ any positive integer, the $(N+1)$-dimensional irreducible representation is given by
\begin{equation}
L_0=\left(\begin{array}{ccccc}
\frac{N}{2} & & & & \\ & \frac{N}{2}-1 & & & \\ & & \ddots & & \\ & & & -\frac{N}{2}+1 & \\ & & & & -\frac{N}{2} \end{array}\right), \quad
L_+= \left(\begin{array}{ccccc}
0 & \sqrt{1\cdot N}& & & \\ & 0 & \sqrt{2\cdot(N-1)}& & \\ & & \ddots &\ddots & \\ & & & & \sqrt{N\cdot 1}\\ & & & & 0 \end{array}\right), 
\label{L0L+}
\end{equation}
and $L_- = (L_+)^T$.
Note that, in this representation, the matrix $M_K$ from~\eqref{MK} 
is written as
\begin{equation}
M_K= \frac{N}{2} I + (2p-1) L_0 -\sqrt{p(1-p)} \left( L_+ + L_-\right),
\end{equation}
where $I$ is the identity matrix. 
Consequently, the computation of
\begin{equation}
f_{r,s}(t)= (r | \exp(-itM_K) |s)
\end{equation}
simply leads to the computation of a matrix element of an $SU(2)$ group 
element. 
So apart from a factor $e^{-itN/2}$, we need the computation of the following 
left hand side, which we write as
\begin{equation}
\exp(-it(2p-1) L_0 +it\sqrt{p(1-p)} \left( L_+ + L_-\right)) = e^{\xi L_-} e^{\eta L_0} e^{\zeta L_+}.
\label{decomp}
\end{equation}
The equation~\eqref{decomp} is an example of the standard 
Baker-Campbell-Hausdorff (BCH) type of decomposition of an $SU(2)$ group 
element. For a general discussion on BCH decomposition of group elements see~\cite{ZFG90}. 
As we intend to use such decompositions in other contexts, 
we provide here the usual procedure of derivation of these rules.   
The constants $\xi,\eta,\zeta$ in~\eqref{decomp} should be representation 
independent (all elements are group elements), so
to determine these we can perform the computation using any faithful 
representation, and, in particular, the standard 2-dimensional representation.
Expanding the exponential in the left hand side of~\eqref{decomp}, using the $2\times 2$ matrices
\begin{equation}
L_0 \rightarrow \left(\begin{array}{cc} 1/2&0\\0&-1/2\end{array}\right),\quad
L_+ \rightarrow \left(\begin{array}{cc} 0&1\\0&0\end{array}\right),\quad
L_- \rightarrow \left(\begin{array}{cc} 0&0\\1&0\end{array}\right),
\label{2drep}
\end{equation} 
one obtains after some calculations:
\begin{equation}
\left( \begin{array}{cc}
\cos\frac{t}{2} -i(2p-1)\sin(\frac{t}{2}) &  2i\sqrt{p(1-p)} \sin(\frac{t}{2}) \\
2i\sqrt{p(1-p)} \sin(\frac{t}{2}) & \cos(\frac{t}{2}) +i(2p-1)\sin(\frac{t}{2})
\end{array}\right).
\end{equation}
On the other hand, the right hand side of~\eqref{decomp} yields, in the same 2-dimensional 
representation~\eqref{2drep}:
\begin{equation}
\left( \begin{array}{cc}
e^{\eta/2} &  \zeta e^{\eta/2} \\
\xi e^{\eta/2} & \xi\zeta e^{\eta/2} + e^{-\eta/2}
\end{array}\right).
\end{equation}
Identification yields:
\begin{equation}
e^{\eta/2} = \cos\bigl(\frac{t}{2}\bigr) -i(2p-1)\sin\bigl(\frac{t}{2}\bigr), \qquad
\xi=\zeta=\frac{2i\sqrt{p(1-p)} \sin(\frac{t}{2})}{\cos(\frac{t}{2}) -i(2p-1)\sin(\frac{t}{2})}.
\label{xi}
\end{equation}
Now we compute the matrix element of the right hand side of~\eqref{decomp} for an arbitrary
representation of dimension $(N+1)$. Using the common matrix elements~\eqref{L0L+}, one finds
\begin{equation}
L_0|s) = (\frac{N}{2}-s)\; |s), \qquad
\frac{(L_+)^k}{k!} |s) = \sqrt{\binom{s}{k}\binom{N-s+k}{k}}\; |s-k),
\end{equation}
and similarly
\begin{equation}
(r| \frac{(L_-)^j}{j!} = \sqrt{\binom{r}{j}\binom{N-r+j}{j}} (r-j|.
\end{equation}
Now (using $\xi=\zeta$)
\begin{align}
&(r| e^{\xi L_-} e^{\eta L_0} e^{\zeta L_+} |s) = 
 \sum_{j,k=0}^N \frac{\xi^{j+k}}{j!k!} (r| (L_-)^j e^{\eta L_0} (L_+)^k |s) \nn\\
& = \sum_{j,k} \xi^{j+k} \sqrt{\binom{r}{j}\binom{N-r+j}{j}\binom{s}{k}\binom{N-s+k}{k}} e^{\eta(\frac{N}{2}-s+k)} 
\delta_{r-j,s-k} \nn\\
& = \sum_j \sqrt{\binom{r}{j}\binom{N-r+j}{j}\binom{s}{s-r+j}\binom{N-r+j}{s-r+j}} 
\xi^{s-r+2j} e^{\eta(\frac{N}{2}-r+j)} \nn\\
& = \sqrt{\frac{r!s!}{(N-r)!(N-s)!}} \xi^{s-r} e^{\eta(\frac{N}{2}-r)} 
\sum_j \frac{(N-r+j)!}{j!(r-j)!(j+s-r)!} (\xi^2 e^{\eta})^j.
\end{align}
The last sum is of hypergeometric type and is proportional to
\begin{equation}
{}_2F_1 \left(  \atop{-r,-s}{-N} ; \frac{-1}{\xi^2 e^\eta} \right)=
{}_2F_1 \left(  \atop{-r,-s}{-N} ; \frac{1}{4p(1-p)\sin^2(\frac{t}{2})} \right),
\end{equation}
using~\eqref{xi}. But $z=e^{-it}$, so $(1-z)^2/z=-4\sin^2(\frac{t}{2})$, and taking all factors
together one recovers~\eqref{frsK}.
Note that this computation is essentially equivalent to that relating matrix elements of finite rotations
($d$-functions) to Jacobi polynomials~\cite[Chapter~4]{Edmonds}.

\subsection{Discussion and special cases}

Let us return to the closed form expression~\eqref{frsK} for the transition amplitude $f_{r,s}(t)$
at time~$t$. As already noticed, since $f_{r,s}(t)$ is a function of $z=e^{-it}$, the system is periodic
with period $2\pi$. At the initial moment $t=0$ the system is in the state with all spins down except at
site $s$ where the spin is up. At any time which is a multiple of $2\pi$, the system is back in this
initial state: $f_{r,s}(2k\pi)=\delta_{rs}$ for $k\in\Z_+$. At other times, the system is in general 
in a mixed state. Due to the orthogonality of the basis states, one has
\begin{equation*}
\sum_{r=0}^N |f_{r,s}(t)|^2 = 1
\end{equation*}
for any $s$ and any time $t$. In fact, more generally, the matrix of correlation functions is
unitary, so for any time $t$ and any indices $r$ and $s$:
\begin{equation*}
\sum_{k=0}^N f^*_{k,r}(t) f_{k,s}(t) = \sum_{k=0}^N f^*_{r,k}(t) f_{s,k} =\delta_{r,s}.
\end{equation*}

Let us consider the case when the sender is located at site~$0$, i.e.\ $s=0$. Then~\eqref{frsK} yields
\begin{equation}
f_{r,0}(t) = \sqrt{\binom{N}{r}}(\sqrt{p(1-p)})^{r} (1-z)^{r}(1-p+pz)^{N-r}.
\end{equation}
So far, $p$ ($0<p<1$) is still a free parameter. A special case occurs when $p=1/2$:
\begin{equation}
f_{r,0}(t) = \frac{1}{2^N}\sqrt{\binom{N}{r}} (1-z)^{r}(1+z)^{N-r}. \qquad (p=1/2)
\end{equation}
Using $z=e^{-it}$, this gives
\begin{equation}
|f_{r,0}(t)| = \sqrt{\binom{N}{r}}\; \left|\sin\bigl(\frac{t}{2}\bigr)\right|^{r}\, \left|\cos\bigl(\frac{t}{2}\bigr)\right|^{N-r}. \qquad (p=1/2)
\end{equation}
In other words,
\begin{equation}
f_{r,0}(\pi) = \delta_{r,N}. \qquad (p=1/2)
\end{equation}
This is the situation of ``perfect state transfer'' described already in~\cite{Albanese2004}: at time
$t=\pi$ the system is in the state with all spins down except at site~$N$ the spin is up. 
So for this time there is perfect state transfer from site~$0$ to site~$N$.

Let us mention here that the condition for perfect state transfer can be deduced from the
corresponding Jacobi matrix itself~\cite{Kay2009}. In order to allow perfect state transfer,
the matrix~\eqref{Ham-M} should be mirror-periodic, i.e.\ $h_n=h_{N-n}$ and $J_n=J_{N-1-n}$ for
all~$n$. Clearly, for~\eqref{MK} this is the case when $p=1/2$, see~\eqref{Jh-K}.

More generally, let us specialize the expression~\eqref{frsK} for time $t=\pi$:
\begin{equation}
f_{r,s}(\pi)= \sqrt{\binom{N}{r}\binom{N}{s}}(\sqrt{p(1-p)})^{r+s} 2^{r+s}(1-2p)^{N-r-s}{\ }_2F_1\left(
 \atop{-r,-s}{-N} ; \frac{1}{4p(1-p)} \right).
\end{equation}
This expression shows once again that taking the free parameter $p=1/2$ yields a special case:
\begin{equation*}
f_{r,s}(\pi) = \delta_{r+s,N}. \qquad (p=1/2)
\end{equation*}
So for $p=1/2$ there is also perfect state transfer between the sites $s$ and $N-s$.

\subsection{A limiting case}
\label{Krawt_lim}
A classical limit of Krawtchouk polynomials are Charlier polynomials. Putting the parameter $p=\alpha/N$,
and letting $N$ go to $+\infty$ yields Charlier polynomials $C_n(x;\alpha)$~\cite{Koekoek}:
\begin{equation}
\lim_{N\rightarrow +\infty} K_n(x;\frac{\alpha}{N},N) = C_n(x;\alpha) = {\ }_2F_0\left(
 \atop{-n,-x}{-} ; -\frac{1}{\alpha} \right),
\end{equation} 
satisfying the orthogonality relations
\begin{equation}
\sum_{x=0}^\infty \frac{\alpha^x}{x!} e^{-\alpha} C_m(x;\alpha) C_n(x;\alpha) = \frac{n!}{\alpha^n} \delta_{mn},
\end{equation}
where $\alpha>0$ is a positive parameter.
The recurrence relation reads
\begin{equation}
x C_n(x;\alpha) = -\alpha C_{n+1}(x;\alpha) + 
(n+\alpha) C_n(x;\alpha) - n C_{n-1}(x;\alpha).
\end{equation}

To see which spin chain corresponds to this limit, we take the appropriate limit in~\eqref{Jh-K}, and 
find for~\eqref{Ham2}:
\begin{equation}
\hatH= \sum_{k=0}^\infty (\alpha+k) a^\dagger_k a_k - \sum_{k=0}^{\infty} \sqrt{\alpha(k+1)}(a_k^\dagger a_{k+1}+a_{k+1}^\dagger a_k).
\label{Ham-charlier}
\end{equation}
The above infinite chain of fermions with nearest-neighbour interaction 
immediately provides a unitary representation of the (central extension of the) Heisenberg algebra 
$\mathfrak{h}_{4}(\mathfrak{a}, \mathfrak{a}^{\dagger}, \mathfrak{N},
\mathfrak{I})$, where the algebraic generators may be constructed as 
\begin{equation}
\mathfrak{a}= \sum_{k=0}^\infty\sqrt{k+1}\; a_k^\dagger a_{k+1},\;\;\;
\mathfrak{a}^\dagger = \sum_{k=0}^\infty\sqrt{k+1}\; a_{k+1}^\dagger a_k,\;\;\;
\mathfrak{N}= \sum_{k=0}^\infty k\; a_k^\dagger a_k,\;\;\;
\mathfrak{I}= \sum_{k=0}^\infty a_k^\dagger a_k.
\label{gen_def}
\end{equation}
Employing the defining anticommutation relations of the fermionic variables~\eqref{anticomm} 
we observe that the above operators satisfy the commutation relations of the
Heisenberg algebra $\mathfrak{h}_{4}(\mathfrak{a}, 
\mathfrak{a}^{\dagger}, \mathfrak{N},\mathfrak{I})$:
\begin{equation}
[\mathfrak{a}, \mathfrak{a}^{\dagger}] = \mathfrak{I},\qquad
[\mathfrak{N}, \mathfrak{a}] = - \mathfrak{a},\qquad
[\mathfrak{N}, \mathfrak{a}^{\dagger}] = \mathfrak{a}^{\dagger}\qquad
[\mathfrak{X}, \mathfrak{I}] =  0 \;\;\;\hbox{where}\;\;\; 
\mathfrak{X} \in \{\mathfrak{N}, \mathfrak{a}, \mathfrak{a}^{\dagger}\},
\label{H4}
\end{equation}
where the unitary representation in terms of the single spin excitation states~\eqref{inverse} reads
\begin{equation}
\mathfrak{a}^{\dagger}\; |k) = \sqrt{k + 1}\; |k + 1),\quad
\mathfrak{a}\; |k) = \sqrt{k}\; |k - 1),\quad
\mathfrak{N}\; |k) = k \;|k).
\label{Fock}
\end{equation}
Using the generators introduced in~\eqref{gen_def} we may express the  
Hamiltonian~\eqref{Ham-charlier} as 
\begin{equation}
\hatH = \mathfrak{N} + \alpha \mathfrak{I} - \sqrt{\alpha} (\mathfrak{a} + 
\mathfrak{a}^{\dagger}).
\label{H_Charlie}
\end{equation}
Following~\cite{ZFG90} the time evolution operator introduced 
in~\eqref{evolution} may now be expressed in the normal ordered 
BCH-factorized form:
\begin{equation}
\mathcal{U}(t) = \exp(\alpha (z - 1))\;\;
\exp(\sqrt{\alpha} (1 - z) \mathfrak{a}^{\dagger})\;\;z^{\mathfrak{N}}\;\;
\exp(\sqrt{\alpha} (1 - z) \mathfrak{a}).
\label{U_BCH}
\end{equation}
It is worth mentioning that the correlation function~\eqref{frs} obtained via 
the operator factorization~\eqref{U_BCH}, and the appropriate limiting value 
easily computed from~\eqref{frsK} precisely agree. We quote the final answer:
\begin{equation}
f_{r,s}(t) = \sqrt{\frac{\alpha^{r+s}}{r!s!}}(1-z)^{r+s} e^{-\alpha+\alpha z} {\ }_2F_0\left(
 \atop{-r,-s}{-} ; \frac{z}{\alpha(1-z)^2} \right).
\label{K_frs_lim}
\end{equation}
Intuitively the above result is easily understood from the well-known~\cite{ZFG90} 
result that the large spin $(\frac{N}{2})$ contraction limit of 
the $su(2)$ algebra~\eqref{sl2} is given by the Heisenberg algebra.   
{}From~\eqref{K_frs_lim} it follows that the state with spin up at position 
$s=0$ emanates over the infinite chain with the correlation function 
given below
\begin{equation}
f_{r,0}(t) = \sqrt{\frac{\alpha^{r}}{r!}}(1-e^{-it})^{r} 
e^{-\alpha+\alpha \exp(-it)}.
\end{equation}
Starting at time $t=0$, some reflection takes place at the ``infinite end'' of the chain, and by
time $t=2\pi$ the system is back in its original state. In the middle of this, at time $t=\pi$,
the state is ``spread'' over the infinite chain according to the amplitude
\begin{equation*}
f_{r,0}(\pi) = e^{-2\alpha}\; \sqrt{\frac{(4\alpha)^r}{r!}}.
\end{equation*}
Note that $\sum_{r=0}^\infty f^2_{r,0}(\pi) =1$, as it should be. The asymptotic
limiting value of the correlation function obtained above may be understood 
in the sense of the leading term in a $\frac{1}{N}$ expansion of the said 
quantity \eqref{frsK} for the fixed $N$ case. It should be interesting to 
obtain successive correction terms, and their group theoretic interpretations, 
to the leading value in the large $N$ limit.

\section{The Jacobi matrix of Hahn polynomials}

\subsection{General correlation function and special cases}

The method outlined in the beginning of the previous section is clear, and this analysis can in principle
be made for any set of discrete orthogonal polynomials.
So in this section we shall consider the more general class of Hahn polynomials 
$Q_n(x;\alpha,\beta,N)$~\cite{Koekoek,Suslov}, characterized by a 
positive integer parameter $N$ and two real parameters $\alpha$ and $\beta$ 
(for orthogonality, one should have $\alpha>-1$ and $\beta>-1$, or $\alpha<-N$ and $\beta<-N$).
The Hahn polynomial of degree $n$ ($n=0,1,\ldots,N$) in the variable $x$ is defined by:
\begin{equation}
Q_n(x) \equiv Q_n(x;\alpha,\beta,N) = {\;}_3F_2 \left( \atop{-n,n+\alpha+\beta+1,-x}{\alpha+1,-N} ; 1 \right).
\label{defQ}
\end{equation}
The orthogonality relation reads:
\begin{equation}
\sum_{x=0}^N w(x) Q_n(x) Q_m(x) = d_n \delta_{mn},
\label{orth-Q}
\end{equation} 
where
\begin{align*}
& w(x) = \binom{\alpha+x}{x} \binom{N+\beta-x}{N-x} \qquad (x=0,1,\ldots,N); \\
& d_n = \frac{n!(N-n)!}{N!^2}\frac{(n+\alpha+\beta+1)_{N+1}(\beta+1)_n}{(2n+\alpha+\beta+1)(\alpha+1)_n}.
\end{align*}
We have used the common notation for hypergeometric series and Pochhammer symbols~\cite{Bailey,Slater}, like
$(a)_n=a(a+1)\cdots(a+n-1)$ for $n=1,2,\ldots$ and $(a)_0=1$; $(a,b,\cdots)_n=(a)_n (b)_n \cdots$, etc.
The three-term recurrence relation is given by:
\begin{equation}
-x Q_n(x)  =  A_n \, Q_{n+1}(x)- (A_n+C_n) \, Q_n(x)
+ C_n \, Q_{n-1}(x), \label{Q-rec1}
\end{equation}
where
\[
A_n= \frac{(n+\alpha+\beta+1)(n+\alpha+1)(N-n)}{(2n+\alpha+\beta+1)(2n+\alpha+\beta+2)}, \quad
C_n=\frac{n(n+\alpha+\beta+N+1)(n+\beta)}{(2n+\alpha+\beta)(2n+\alpha+\beta+1)}.
\]
Introducing orthonormal Hahn functions
\begin{equation}
\tilde Q_n(x) \equiv \frac{\sqrt{w(x)} Q_n(x)}{\sqrt{d_n}}
\end{equation}
one has the following result~\cite{Regniers2009}:
\begin{lemm}
Let $M_Q$ be the tridiagonal $(N+1)\times(N+1)$-matrix (Jacobi matrix)
\begin{equation}
\label{MQ}
M_Q= \left( \begin{array}{ccccc}
             h_0 & -J_0  &    0   &        &      \\
            -J_0 &  h_1  &  -J_1  & \ddots &      \\
              0  & -J_1  &   h_2  & \ddots &  0   \\
                 &\ddots & \ddots & \ddots & -J_{N-1} \\
                 &       &    0   &  -J_{N-1}  &  h_N
          \end{array} \right)
\end{equation}
where
\begin{align} 
J_n &=\sqrt{ \frac{(n+1) \, (n+\alpha+1) \, (n+\beta+1) \, (n+\alpha+\beta+1) \, (n+\alpha+\beta+N+2) \, (N-n)}
                  {(2n+\alpha+\beta+2)^2(2n+\alpha+\beta+1)(2n+\alpha+\beta+3)} },\nn\\
h_n &= \frac{N}{2} + \frac{(\alpha-\beta) \bigl[ (\alpha+\beta) \, (N-2n) - 2n(n+1) \bigr]}
                         {2(2n+\alpha+\beta) \, (2n+\alpha+\beta+2)}.
\label{Q_Jh}
\end{align}
and let $U$ be the $(N+1)\times(N+1)$-matrix with elements $U_{jk}=\tilde Q_j(k)$.
Then 
\begin{equation}
U U^T = U^TU=I \qquad\hbox{and}\qquad M_Q=UDU^T
\end{equation}
where
\begin{equation}
D= \diag (0,1,2,\ldots,N).
\end{equation}
\end{lemm}
So for a system corresponding to the quantities~\eqref{Q_Jh}, the transition amplitude is given by
\begin{equation}
f_{r,s}(t) = \frac{1}{\sqrt{d_r d_s}} \sum_{k=0}^N w(k) Q_r(k) Q_s(k) z^k \qquad (z=e^{-it}).
\label{sumQ}
\end{equation}
The purpose is now to compute~\eqref{sumQ}, and then to investigate some special cases.
Let us denote the summation in~\eqref{sumQ} by $S(r,s)$:
\begin{equation}
S(r,s) = \sum_{k=0}^N w(k) Q_r(k) Q_s(k) z^k.
\label{S}
\end{equation}
In order to perform this summation, one can use the following product formula for Hahn polynomials:
\begin{align}
&Q_r(k) Q_s(k)= {\;}_3F_2 \left( \atop{-k,-r,r+\alpha+\beta+1}{-N,\alpha+1} ; 1 \right)
{\;}_3F_2 \left( \atop{-k,-s,s+\alpha+\beta+1}{-N,\alpha+1} ; 1 \right) \nn\\
&= \frac{(-N-\beta)_k}{(\alpha+1)_k} \sum_{m=0}^k
 \frac{(-k,r-N,s-N,-r-c,-s-c)_m}{(1,-N-\beta,-c,-N,-N)_m} \nn \\
& \times  {\;}_8F_7 \left( \atop{c-m,1+\frac{c-m}{2},N+\beta+1-m,-m,-r,-s,c+r-N,c+s-N }
{\frac{c-m}{2}, \alpha+1,c+1,c+1+r-m,c+1+s-m,N+1-r-m,N+1-s-m} ; -1 \right)
\label{8F7}
\end{align}
where we have used the abbreviation $c=N+1+\alpha+\beta$.
This expression can be obtained from the product formula for $q$-Racah polynomials given 
in~\cite[eq.~(8.3.1)]{Gasper}: in this formula, first take the limit $a\rightarrow 0$, and then
take the limit $q\rightarrow 1$.\\
Now we multiply the right hand side of~\eqref{8F7} by $w(k) z^k$, and sum over $k$ from $0$ to $N$. 
Changing the order of summation (over $k$ and $m$), the inner sum over $k$ can be performed using
the binomial theorem. This leads to:
\begin{align}
S(r,s)& = \frac{(\beta+1)_N}{N!}\sum_{m=0}^N (-z)^m(1-z)^{N-m} {\;}_8F_7(-1) \nn\\
&\times \frac{(r-N,s-N,-r-N-\alpha-\beta-1,-s-N-\alpha-\beta-1)_m}{m!(-N,-N-\beta,-N-\alpha-\beta-1)m} ,
\label{Sexplicit}
\end{align}
where the ${\;}_8F_7(-1)$ has the same parameters as in~\eqref{8F7}. 
So together with the factor $1/\sqrt{d_rd_s}$ in~\eqref{sumQ}, \eqref{Sexplicit} gives us a symmetric
and compact formula for the computation of $f_{r,s}(t)$ in the Hahn case.

Let us consider the special case when the sender is at one end of the chain, i.e.~$s=0$. 
Then~\eqref{Sexplicit} gives
\begin{align}
S(r,0)&= \frac{(\beta+1)_N}{N!}\sum_m (-z)^m (1-z)^{N-m} \frac{(r-N,-r-N-\alpha-\beta-1)_m}{m!(-N-\beta)m}\nn\\
&= \frac{(\beta+1)_N}{N!}(1-z)^{N} {\;}_2F_1 \left( \atop{r-N,-r-N-\alpha-\beta-1}{-N-\beta} ; 
 \frac{z}{z-1} \right) \nn\\
&=  \frac{(\beta+1)_N}{N!}(1-z)^{r} {\;}_2F_1 \left( \atop{r-N,r+\alpha+1}{-N-\beta} ; z \right).
\label{fr0}
\end{align}
In the last step, Euler's transformation~\cite{Bailey,Slater} formula was used.
Thus, the transition amplitude becomes
\begin{align}
f_{r,0}(t) &= \left( \binom{N}{r} \frac{(2r+\alpha+\beta+1) (\alpha+1)_r}{(\beta+1)_r (\alpha+\beta+2)_N
(r+\alpha+\beta+1)_{N+1}}\right)^{1/2} \nn\\
&\times (\beta+1)_N (1-z)^r {\;}_2F_1 \left( \atop{r-N,r+\alpha+1}{-N-\beta} ; z \right).
\end{align}
In particular,
\begin{equation}
f_{N,0}(t) = \left( \frac{ (\alpha+1,\beta+1)_N}{(\alpha+\beta+2)_N
(N+\alpha+\beta+1)_{N}}\right)^{1/2} (1-z)^N,
\label{fN0}
\end{equation}
and
\begin{equation}
|f_{N,0}(t)| = \left( \frac{ (\alpha+1,\beta+1)_N}{(\alpha+\beta+2)_N
(N+\alpha+\beta+1)_{N}}\right)^{1/2} 2^N \left|\sin\bigl(\frac{t}{2}\bigr)\right|^N.
\label{fN02}
\end{equation}
An interesting special case is that with the parameters $\alpha$ and $\beta$ equal, because then the
magnetic field strengths $h_k$ in~\eqref{Ham2} are all constant (independent of~$k$), see~\eqref{Q_Jh}.
For $\beta=\alpha$, \eqref{fN02} becomes
\begin{equation}
|f_{N,0}(t)| = \left( \frac{ (\alpha+1)_N }{(\alpha+3/2)_{N-1}\,
(\alpha+N/2+1/2)}\right)^{1/2}  \left|\sin\bigl(\frac{t}{2}\bigr)\right|^N.
\label{fN03}
\end{equation}
Clearly, this is maximal for $t=\pi$ (plus multiples of $2\pi$). However, $|f_{N,0}(t)|<1$ for
the allowed values of $\alpha$. Only for large $\alpha$, $|f_{N,0}(t)|$ approaches~$1$. 
So ``perfect state transfer'' between site 0 and site $N$ does not take place except for
$\alpha\rightarrow +\infty$. This limiting case does not give rise to a new example:
for $\alpha=\beta \rightarrow +\infty$, the Hahn polynomials reduce to Krawtchouk polynomials
with $p=1/2$, and this was the subject of the previous section.

\subsection{A limiting case}

A classical limit of Hahn polynomials are Meixner polynomials. Putting $\alpha=b-1$, $\beta=N\frac{1-c}{c}$,
and letting $N$ go to $+\infty$ yields Meixner polynomials $M_n(x;b,c)$~\cite{Koekoek}:
\begin{equation}
\lim_{N\rightarrow +\infty} Q_n(x;b-1,N\frac{1-c}{c},N) = M_n(x;b,c) = {\ }_2F_1\left(
 \atop{-n,-x}{b} ; 1-\frac{1}{c} \right),
\end{equation} 
satisfying the orthogonality relations
\begin{equation}
\sum_{x=0}^\infty \frac{(b)_x}{x!} c^{x} M_m(x;b,c) M_n(x;b,c) = \frac{c^{-n}n!}{(b)_n(1-c)^b}\; \delta_{mn},
\end{equation}
where $b>0$ is a positive parameter and $0<c<1$.
For the recurrence relation, see~\cite{Koekoek}.

Again one can wonder which spin chain corresponds to this limit. Taking the appropriate limits in~\eqref{Q_Jh}, 
one finds for~\eqref{Ham2}:
\begin{equation}
\hatH= \sum_{k=0}^\infty \frac{k+c(k+b)}{1-c} a^\dagger_k a_k -
\sum_{k=0}^{\infty} \frac{\sqrt{c(k+1)(k+b)}}{1-c}
(a_k^\dagger a_{k+1}+a_{k+1}^\dagger a_k). 
\label{Ham-meixner}
\end{equation}
It may be immediately observed that the $su(1,1)$ algebra acts as the spectrum 
generating algebra of the above Hamiltonian. We define the $su(1,1)$
generators and the identity operator as bilinear constructs of the fermionic 
variables by
\begin{align}
\mathcal{K}_{0} &= \sum_{k=0}^\infty (k + \frac{b}{2})\;a^\dagger_k a_k,\nn\\
\mathcal{K}_{+} &= \sum_{k=0}^\infty \sqrt{(k+1)(k+b)}\;a_{k+1}^\dagger a_k,
\nn\\
\mathcal{K}_{-} &= \sum_{k=0}^\infty \sqrt{(k+1)(k+b)}\;a_{k}^\dagger a_{k+1},
\nn\\
\mathcal{I} &= \sum_{k=0}^\infty a^\dagger_k a_k.
\label{su11_gen}
\end{align}
By virtue of~\eqref{anticomm} the above generators satisfy the $su(1,1)$ 
commutation relations:
\begin{equation}
[\mathcal{K}_{0}, \mathcal{K}_{\pm}] = \pm \mathcal{K}_{\pm},\qquad
[\mathcal{K}_{+}, \mathcal{K}_{-}] = - 2 \mathcal{K}_{0},\qquad
[\mathcal{X}, \mathcal{I}] = 0 \;\;\;\hbox{where}\;\;\;
\mathcal{X} \in \{\mathcal{K}_{0},\; \mathcal{K}_{\pm}\}.
\label{su11_alg}
\end{equation}
The large $N$ Hamiltonian~\eqref{Ham-meixner} now assumes the form  
\begin{equation}
\hat{H} = \frac{1 + c} {1 - c}\;\mathcal{K}_{0} - \frac{b}{2} \mathcal{I} 
-\frac{\sqrt{c}}{1 - c} \;(\mathcal{K}_{+} + \mathcal{K}_{-}).
\label{H_su11} 
\end{equation}
The infinite-dimensional lowest weight representation of $su(1,1)$ reads 
(see, e.g.~\cite[eq.~(2.2)]{VdJ1999} or~\cite{Bacry})
\begin{align}
\mathcal{K}_{+}\; \Big|\frac{b}{2},n\Big) &= 
\sqrt{(n + 1)\;(n + b)}\;\Big|\frac{b}{2},n+1\Big),\nn\\ 
\mathcal{K}_{-}\; \Big|\frac{b}{2},n\Big) &= 
\sqrt{n \;(n + b - 1)}\;\Big|\frac{b}{2},n-1\Big),\nn\\
\mathcal{K}_{0}\; \Big|\frac{b}{2},n\Big) &= \Big(n + \frac{b}{2}\Big)\;
\Big|\frac{b}{2},n\Big),
\label{su11_rep}
\end{align}
where the lowest weight $\frac{b}{2}$ has been made explicit in the notation  
of the state vector.
Following~\cite{ZFG90} the BCH-factorization of the time evolution operator~\eqref{evolution} 
may now be easily obtained:
\begin{equation}
\mathcal{U}(t) = \frac{1}{\sqrt{z^{b}}}\;\;
\exp \Big(\sqrt{c}\;\frac{1 - z}{1 - c z}\;\mathcal{K}_{+}\Big)\;\;
\left(\frac{(1 - c) \sqrt{z}}{1 - c z}\right)^{2 \mathcal{K}_{0}} 
\;\;\exp \Big(\sqrt{c}\;\frac{1 - z}{1 - c z}\;\mathcal{K}_{-}\Big).
\label{su11_BCH}
\end{equation}
Employing the decomposition \eqref{su11_BCH} the correlation function~\eqref{frs} 
for the asymptotic limit of the spin chain governed by the 
Hamiltonian~\eqref{H_su11} is readily obtained. As expected, this precisely 
agrees with the appropriate limit that
can easily be computed from~\eqref{Sexplicit}:
\begin{equation}
f_{r,s}(t) = (1-c)^b \sqrt{\frac{(b)_r(b)_s}{r!s!}} c^{(r+s)/2} \frac{(1-z)^{r+s}}{(1-cz)^{b+r+s}}
{\ }_2F_1\left( \atop{-r,-s}{b} ; c\Big(1-\frac{1}{c}\Big)^2\frac{z}{(1-z)^2} 
\right).
\end{equation}
As noted in section~\ref{Krawt_lim} the above asymptotic limit of the 
transition amplitude~\eqref{sumQ} may be understood as its leading 
term in a $\frac{1}{N}$ 
expansion scheme. Let us consider an example here, say for $b=1$ and $c=1/2$.
At time $t=0$ the state with spin up at position $s=0$ is ``released'' over 
the infinite chain;
then at time $t=\pi$ it is ``spread'' as follows:
\begin{equation}
f_{r,0}(\pi) = \frac{1}{3} \left( \frac{\sqrt{8}}{3} \right)^r ;
\end{equation}
(verify that $\sum_{r=0}^\infty f^2_{r,0}(\pi)=1$).
So it decays exponentially over the chain, only to return back to its original configuration
at time $t=2\pi$.

\section{The Jacobi matrix of dual Hahn and Racah polynomials}

\subsection{General computation for dual Hahn polynomials}

Dual Hahn polynomials will play a special role.
First of all, the energy spectrum is not linear. Secondly, under certain conditions they will
allow perfect state transfer.

Dual Hahn polynomials $R_n(\lambda(x);\gamma,\delta,N)$~\cite{Koekoek,Suslov} are characterized by a 
positive integer parameter $N$ and two real parameters $\gamma$ and $\delta$ 
(for orthogonality, one should have $\gamma>-1$ and $\delta>-1$, or $\gamma<-N$ and $\delta<-N$).
The dual Hahn polynomial is not a polynomial of degree $n$ in $x$, but of degree $n$ ($n=0,1,\ldots,N$) 
in $\lambda(x)=x(x+\gamma+\delta+1)$:
\begin{equation}
R_n(\lambda(x)) \equiv R_n(\lambda(x);\gamma,\delta,N) 
= {\;}_3F_2 \left( \atop{-n,-x,x+\gamma+\delta+1}{\gamma+1,-N} ; 1 \right).
\label{defR}
\end{equation}
The orthogonality relation reads:
\begin{equation}
\sum_{x=0}^N w(x) R_n(\lambda(x)) R_m(\lambda(x)) = d_n \delta_{mn},
\label{orth-R}
\end{equation} 
where
\begin{align*}
& w(x) = \frac{(2x+\gamma+\delta+1)(\gamma+1)_x (-N)_x N!}{(-1)^x(x+\gamma+\delta+1)_{N+1}(\delta+1)_x x!} 
\qquad (x=0,1,\ldots,N); \\
& d_n^{-1} = \binom{\gamma+n}{n} \binom{\delta+N-n}{N-n}.
\label{w-h-R}
\end{align*}
The three-term recurrence relation is given by:
\begin{equation}
\lambda(x) R_n(\lambda(x))  =  A_n \, R_{n+1}(\lambda(x))- (A_n+C_n) \, R_n(\lambda(x))
+ C_n \, R_{n-1}(\lambda(x)), 
\label{R-rec1}
\end{equation}
where
\[
A_n= (n+\gamma+1)(n-N), \qquad
C_n=n(n-\delta-N-1).
\]
Orthonormal dual Hahn functions are defined by
\begin{equation}
\tilde R_n(\lambda(x)) \equiv \frac{\sqrt{w(x)} R_n(\lambda(x))}{\sqrt{d_n}},
\end{equation}
and then one can deduce:
\begin{lemm}
Let $M_R$ be the tridiagonal $(N+1)\times(N+1)$-matrix (Jacobi matrix)
\begin{equation}
\label{MR}
M_R= \left( \begin{array}{ccccc}
             h_0 & -J_0  &    0   &        &      \\
            -J_0 &  h_1  &  -J_1  & \ddots &      \\
              0  & -J_1  &   h_2  & \ddots &  0   \\
                 &\ddots & \ddots & \ddots & -J_{N-1} \\
                 &       &    0   &  -J_{N-1}  &  h_N
          \end{array} \right)
\end{equation}
where
\begin{align} 
J_n &=\sqrt{ (n+1) \, (n+\gamma+1) \,(N-n)\,(\delta+N-n)  },\nn\\
h_n &= (n+\gamma+1)(N-n)+n(\delta+N-n+1).
\label{R_Jh}
\end{align}
and let $U$ be the $(N+1)\times(N+1)$-matrix with elements $U_{jk}=\tilde R_j(\lambda(k))$.
Then 
\begin{equation}
U U^T = U^TU=I \qquad\hbox{and}\qquad M_R=UDU^T
\end{equation}
where
\begin{equation}
D= \diag (\epsilon_0,\epsilon_1,\epsilon_2,\ldots,\epsilon_N) \hbox{ with } \epsilon_k=k(k+\gamma+\delta+1).
\label{R-D}
\end{equation}
\end{lemm}
Note that due to the appearance of $\lambda(x)$ in~\eqref{R-rec1}, the matrix~$D$ has the form~\eqref{R-D},
and thus the energy eigenvalues of the single fermion Hamiltonian eigenstates are quadratic in~$k$.

The rest of the analysis is again concerned with the general correlation function. 
For a system corresponding to the quantities~\eqref{R_Jh}, this is now given by
\begin{equation}
f_{r,s}(t) = \frac{1}{\sqrt{d_r d_s}} \sum_{k=0}^N w(k) R_r(\lambda(k)) R_s(\lambda(k)) 
z^{k(k+\gamma+\delta+1)} \qquad (z=e^{-it}).
\label{sumR}
\end{equation}
In this case, one can use the product formula for $q$-Hahn polynomials~\cite[Eq.~(8.3.3)]{Gasper} and
take the limit $q\rightarrow 1$ to find:
\begin{align}
R_r(\lambda(k)) R_s(\lambda(k))&= {\;}_3F_2 \left( \atop{-r,-k,k+\gamma+\delta+1}{\gamma+1,-N} ; 1 \right)
{\;}_3F_2 \left( \atop{-s,-k,k+\gamma+\delta+1}{\gamma+1,-N} ; 1 \right)  \nn\\
&= (-1)^k \frac{(\delta+1)_k}{(\gamma+1)_k} \sum_{m=0}^k
 \frac{(-k,r-N,s-N,\gamma+\delta+k+1)_m}{(1,\delta+1,-N,-N)_m} \nn\\
& \times {\;}_4F_3 \left( \atop{-m,-r,-s,-\delta-m }
{\gamma+1,N+1-r-m,N+1-s-m} ; 1 \right).
\label{4F3}
\end{align}
Then one obtains, using~\eqref{4F3} and exchanging the order of summation:
\begin{align}
f_{r,s}(t) &= \frac{1}{\sqrt{d_rd_s}} \sum_{m=0}^N 
 \frac{(r-N,s-N)_m}{(\delta+1)_m} 
{\;}_4F_3 \left( \atop{-m,-r,-s,-\delta-m }
{\gamma+1,N+1-r-m,N+1-s-m} ; 1 \right) \nn\\
& \times (-1)^m\frac{((N-m)!)^2}{m!N!} \sum_{k=m}^N \frac{(-N)_k(\gamma+\delta+k+1)_m(\gamma+\delta+2k+1)}{(k-m)!(\gamma+\delta+k+1)_{N+1}}
z^{k(k+\gamma+\delta+1)}.
\label{sum-R}
\end{align}
Due to the appearance of $z^{k(k+\gamma+\delta+1)}$, the inner sum is no longer of hypergeometric type, and it cannot
be simplified in general.
Let us therefore specialize to the case with $s=0$ and $r=N$ (sending site at one end and receiving site at the other
end of the chain).
Then
\begin{equation}
f_{N,0}(t) = \sqrt{(\gamma+1,\delta+1)_N} \;\sum_{k=0}^N 
\frac{(-N)_k (\gamma+\delta+2k+1)}{k!(\gamma+\delta+k+1)_{N+1}}
z^{k(k+\gamma+\delta+1)}.
\label{fN0-R}
\end{equation}
So far, $\gamma$ and $\delta$ are free parameters. 
Let us now require the following condition:
$\gamma+\delta$ is an {\em odd integer number}. 
Then at time $t=\pi$ one has
$z^{k(k+\gamma+\delta+1)}=(-1)^{k(k+\gamma+\delta+1)}=(-1)^k$. 
The summation over $k$ in~\eqref{fN0-R} can now be performed, since it corresponds to
a nearly-poised ${}_3F_2(-1)$ (see~\cite[(III.25)]{Slater}). 
One obtains:
\begin{equation}
f_{N,0}(\pi) = \frac{\sqrt{(\gamma+1,\delta+1)_N}}{(\frac{\gamma+\delta}{2}+1)_N}, 
\qquad(\gamma+\delta=\hbox{ odd integer}).
\label{fN0pi-R}
\end{equation}
Clearly, this expression assumes its maximum value for $\gamma=\delta$, and in that case it is equal to 1.
In other words, for $\gamma=\delta=p+\frac{1}{2}$, with $p$ an integer, there is 
perfect state transfer between the sites $0$ and $N$ at time $t=\pi$.

Similarly, when $\gamma+\delta$ is of the form
\begin{equation}
\gamma+\delta = \frac{2p+1}{q}, \qquad p,q\in\N \qquad(q\ne 0),
\end{equation}
it also follows that at time $t=q\pi$
\begin{equation}
z^{k(k+\gamma+\delta+1)}=e^{-iq\pi [k(k+1)+\frac{2p+1}{q}k]} = e^{-i\pi(2p+1)k}=(-1)^k.
\end{equation}
So in that case $f_{N,0}(q\pi)$ assumes the same value as given by the right hand side of~\eqref{fN0pi-R}.
Thus for $\gamma=\delta=\frac{2p+1}{2q}$, one has perfect state transfer from 0 to $N$ at time $t=q\pi$.
This is the situation described in~\cite{Albanese2004}.

\subsection{The case of Racah polynomials}

Dual Hahn polynomials are a limiting case of Racah polynomials.
Among the discrete orthogonal polynomials, Racah polynomials are the most general, but
also the most complicated.
Their Jacobi matrix is not mirror-periodic, so perfect state transfer is not possible~\cite{Kay2009}.
Racah polynomials $R_n(\lambda(x);\alpha,\beta,\gamma,\delta)$ are polynomials of degree~$n$
in the variable $\lambda(x)=x(x+\gamma+\delta+1)$, and are expressed as a ${}_4F_3(1)$ series,
where one of the numerator parameters $\alpha+1$, $\beta+\delta+1$ or
$\gamma+1$ should be $-N$, with $N$ a positive integer~\cite{Koekoek}. 
Without loss of generality, let us assume we are in the first case with $\alpha+1=-N$.
Then, for the weight function to be positive, assume
\[
\gamma+1>0,\quad \delta+1>0,\quad \beta>\gamma+N.
\]
The orthogonality relation and the coefficients of the recurrence relation become
fairly complicated, see~\cite{Koekoek}. 
Moreover, the single fermion eigenvalues are of the same form as~\eqref{R-D}.
This means that the correlation function is given by~\eqref{sumR}, where $w(k)$ and $d_n$ now
stand for the weight function and squared norm of the Racah polynomials respectively,
and where $R_n(\lambda(k))$ is a Racah polynomial.
Due to the appearance of $z^{k(k+\gamma+\delta+1)}$, the final summation is again
no longer of hypergeometric type and cannot be simplified in general.
Let us therefore not give the general expression, but only some special cases.
For $s=0$ and $r=N$, the correlation function becomes
\begin{equation}
f_{N,0}(t) = \frac{1}{\sqrt{d_Nd_0}} \sum_{k=0}^N \frac{(-N, \gamma+\delta+1, (\gamma+\delta+1)/2+1)_k}{k!(\gamma+\delta+N+2, (\gamma+\delta+1)/2)_k} z^{k(k+\gamma+\delta+1)}.
\label{sum-Racah}
\end{equation}
When $z^{k(k+\gamma+\delta+1)}=(-1)^k$, this sum becomes a nearly-poised ${}_3F_2(-1)$
series which can be summed using~\cite[(III.25)]{Slater}. 
In other words, when $\gamma+\delta$ is an {\em odd integer number}, one finds
\begin{equation}
f_{N,0}(\pi) = \sqrt{\frac{(\gamma+1-\beta,\delta+1+\beta)_N}{(\beta,-\beta)_N}}\; \frac{\sqrt{(\gamma+1,\delta+1)_N}}{(\frac{\gamma+\delta}{2}+1)_N}.
\label{fN0pi-Racah}
\end{equation}
Note that in the limit $\beta\rightarrow +\infty$, in which case the Racah polynomials
become dual Hahn polynomials, \eqref{fN0pi-Racah} indeed becomes~\eqref{fN0pi-R}.

\section{Discussion and conclusion}

In this paper, we have considered linear spin chains with a nearest-neighbour hopping interaction, 
as models for quantum communication. 
We have considered the time evolution of single fermion states in such a spin chain. 
In particular, if the system is at time $t=0$ in a pure state with all spins down except
one spin up at site $s$ of the chain, we have studied the behaviour of this system at time~$t$ 
by computing the transition amplitude $f_{r,s}(t)$.
The main contribution of the paper is to show that one can deduce closed form expressions
for this transition amplitude (or correlation function)
if the interaction matrix of the system is related to the
Jacobi matrix of a set of (discrete) orthogonal polynomials. 

We have worked out in detail the cases related to Krawtchouk polynomials (section~2), 
Hahn polynomials (section~3) and dual Hahn polynomials (section~4); for the case of
Racah polynomials we give only some partial result in section~4.2.
Experts in orthogonal polynomials might wonder why we did not proceed the opposite way, 
starting from the most general case (Racah polynomials), and then obtaining the other
cases as certain limits. 
This approach would work here only for the limit from Racah polynomials to dual Hahn
polynomials, because of the appearance of $z^{k(k+\gamma+\delta+1)}$ in~\eqref{sum-Racah}
and~\eqref{sumR}. For the other correlation functions, there is just $z^k$ in the 
summation part, and these need to be treated separately anyway.
The Krawtchouk case could have been presented as a limit of the Hahn case; however, we
felt it was better to start with a simple example first, which has moreover some additional
interesting properties (such as the group-theoretical interpretation, the 
special case of perfect state transfer, and an interesting limit of its own).
The case of Hahn and dual Hahn polynomials had to be considered separately 
because of the different nature of the correlation function ($z^k$ as opposed
to $z^{k(k+\gamma+\delta+1)}$).

For all examples considered here, we have obtained complete or partial results.
In the case of Krawtchouk polynomials, we obtained a simple 
closed form expression
for $f_{r,s}(t)$ in general. This was also the case for a limit consisting
of an infinite chain of spins described by the Jacobi matrix of Charlier 
polynomials. We noticed that this example is related to the unitary 
representation of the Heisenberg algebra.  
In the case of Hahn polynomials, we do obtain a general 
expression~\eqref{Sexplicit} 
for $f_{r,s}(t)$, though it is still quite complicated. Some special cases
have been discussed, as well as the limit related to Meixner polynomials
(where a simple general expression is obtained). We have related this example 
with the $su(1,1)$ symmetry algebra of the corresponding Hamiltonian. The 
asymptotic $N \rightarrow \infty$ limit of the spin chain may be understood in 
the framework of a $\frac{1}{N}$ expansion where the leading terms of the 
correlation functions of the related Hamiltonians are obtained.   
In the case of dual Hahn polynomials, the general expression for $f_{r,s}(t)$
contains a summation part which is not of hypergeometric type, but we do show
that it simplifies in special cases. The same remarks hold for the case of 
Racah polynomials.

\par

It is worth discussing certain related areas where our analysis may find 
extensions or applications. Propagation of entangled states in anisotropic spin 
chains has been studied in~\cite{AOPFP04}. Anisotropic models are characterized
by the property that they allow instantaneous creation of pairwise 
entanglement from the fully polarized ground state. The anisotropy parameter 
connects~\cite{AOPFP04} the isotropic $XY$ model with the quantum Ising model, 
which, in the $N \rightarrow \infty$ limit, undergoes a quantum 
phase transition at a critical value of the coupling 
constant. In our case it should be of interest to understand the dynamics of 
propagation of entangled states in a spin chain governed by an anisotropic 
variation of the Hamiltonian~\eqref{Ham1}, where the coupling constants follow
the polynomial structures considered here. A class of Hamiltonians that do not 
preserve the total number of excited spins 
was found~\cite{FPTH08} to dynamically create multipartite 
entangled states starting from an initially uncorrelated state. In this 
context one may also introduce  the one-axis spin squeezing interactions in the
Hamiltonian~\cite{KU93} that is expected to protect the entangled states 
against decoherence. The group theoretic method developed in our work may 
help in describing analytical solutions for evolutions of such entangled 
states in the quantum register. Work towards this is in progress.  

\par

We complete our work mentioning another context in which the formalism 
developed here may be relevant. A recent work~\cite{MCHGH09} considers a
Jaynes-Cummings-Hubbard (JCH) system 
that describes coupled cavity structures where confined photons are induced to 
interact via their coupling to embedded two-state systems. In particular, 
a nonuniform ``parabolic'' coupling between the cavities is assumed in~\cite{MCHGH09} 
that is identical to our discussions in section~\ref{Krawtchouk} 
regarding the nearest-neighbour coupling guided by the Krawtchouk polynomials.
The large $N$ asymptotic limit of the correlation function obtained in 
\eqref{K_frs_lim} may be useful in understanding the mean field results, and
consequently, the quantum phase transitions for the JCH Hamiltonians. Moreover,
our examples of Jacobi matrices corresponding to Hahn polynomials and their 
asymptotic limits, and the dual Hahn polynomials may also be considered as 
pre-engineered couplings between the cavities for JCH systems. Our evaluation 
of the correlation functions for these cases may have some importance in
developing the theory of JCH systems with inter-cavity couplings subject to 
these polynomial structures. 

\section*{Acknowledgments}
This research was supported by project P6/02 of the Interuniversity Attraction Poles Programme (Belgian State -- 
Belgian Science Policy).
R.\ Chakrabarti wishes to acknowledge Ghent University for a visitors grant.
J.\ Van der Jeugt wishes to thank M.\ Rahman for pointing out the relevance 
of~\cite[eq.~(8.3.1)]{Gasper}
in this context. We thank E.I.\ Jafarov for discussions and for pointing out 
Ref.~\cite{MCHGH09} to us.

\end{document}